\documentclass[aip,pof,reprint,onecolumn]{revtex4-1} 
\usepackage{graphicx}
\usepackage{dcolumn}
\usepackage{bm}
\usepackage{amssymb}
\usepackage{amsmath}
\usepackage{amsfonts}
\usepackage{mathtools}
\usepackage{comment}
\usepackage{lineno}
\usepackage{soul}
\usepackage{natbib}
\usepackage{color}     
\usepackage{soul}
\usepackage[usenames,dvipsnames,svgnames,table]{xcolor} 
\usepackage[colorlinks=true,
            linkcolor=blue,
            urlcolor=blue,
            citecolor=BrickRed]{hyperref}
            
\usepackage{mathrsfs}  

\usepackage{tabulary}

\usepackage{booktabs}

\usepackage{mathtools}
\usepackage{amssymb}
\usepackage{amsmath}

\usepackage[abs]{overpic}
\usepackage{xcolor,varwidth}   
\definecolor{light-gray}{gray}{0.95}

\usepackage{fancyhdr}

\newcommand{\conferenceheader}{%
  \small\itshape
  Proceedings of the 1st International Symposium on AI and Fluid Mechanics,
  Chania, Greece, 2025%
}

\fancypagestyle{plain}{
  \fancyhf{}
  \fancyhead[C]{\conferenceheader}
  \fancyfoot[C]{\thepage}

}

\begin{document}
\title{Physics-guided Bayesian neural networks for zonal corrections and uncertainty quantification in separated flows}

\author{Ali Eidi}
\affiliation{Department of Flow Physics \& Technology, Faculty of Aerospace Engineering, TU Delft, 2629 HS Delft, The Netherlands}
\author{Tyler Buchanan}
\affiliation{Department of Flow Physics \& Technology, Faculty of Aerospace Engineering, TU Delft, 2629 HS Delft, The Netherlands}
\author{Letian Jiang}
\affiliation{Department of Flow Physics \& Technology, Faculty of Aerospace Engineering, TU Delft, 2629 HS Delft, The Netherlands}
\affiliation{Chair of Aerodynamics and Fluid Mechanics, School of Engineering and Design, Technical University of Munich, 85748 Garching bei M\"unchen, Germany}
\author{Richard P.~Dwight}
\email{r.p.dwight@tudelft.nl}
\affiliation{Department of Flow Physics \& Technology, Faculty of Aerospace Engineering, TU Delft, 2629 HS Delft, The Netherlands}

\begin{abstract}
Data-driven techniques have improved the accuracy of Reynolds-averaged Navier-Stokes (RANS) models in fluid dynamics. However, modeling separated flows remains challenging due to their complex physics and sensitivity to local conditions. Existing approaches often struggle with generalization beyond training cases and lack robust uncertainty quantification frameworks, limiting their utility in complex flow regimes. We propose a Bayesian neural network (BNN)-based framework specifically designed for two-dimensional separated flows. By focusing on flow zones near separated regions, we ensure targeted training and enhance predictive reliability. The BNN framework incorporates physics-guided, invariant inputs to maintain consistency with turbulence physics. Correction terms predicted by the BNN are selectively applied to specific regions of the flow domain using a novel classifier, improving accuracy. A key feature of this approach is propagating BNN-derived corrections to flow solutions, enabling uncertainty quantification in unseen test cases. This probabilistic characterization of modeling errors offers insights into the reliability of RANS predictions across geometries with similar topologies. Preliminary results demonstrate that this method accurately predicts correction terms for Reynolds stress anisotropy and turbulent kinetic energy production in separated flow regions, effectively addressing dominant modeling errors and advancing turbulence modeling through uncertainty quantification.
\end{abstract}
\maketitle
\thispagestyle{fancy}
\section{Introduction}

Reynolds-averaged Navier--Stokes (RANS) simulations are widely used for turbulence modeling due to their efficiency but struggle in separated flows, where empirically tuned closures fail to capture complex flow physics. While large-eddy simulation (LES) improves accuracy, its high computational cost limits practical applications, driving the need for enhanced RANS models that balance accuracy and efficiency. Data-driven methods have been explored to refine RANS predictions, from parameter tuning to machine-learning-based Reynolds-stress corrections. Notably, symbolic regression~\cite{schmelzer2020discovery} provides interpretable corrections but is constrained by predefined functional forms, while tensor-basis neural networks~\cite{ling2016reynolds} offer greater flexibility but lack interpretability and are challenging to integrate into CFD solvers.

Moreover, RANS models and most of the existing data-driven approaches provide deterministic corrections without quantifying uncertainty, which is crucial for reliable turbulence predictions. They often overlook epistemic uncertainty, which arises from model limitations and can be reduced with more data, and aleatoric uncertainty, which stems from inherent variability in the data and remains even with additional observations. Accounting for these uncertainties enhances the robustness of RANS simulations, especially in separated flows, where turbulence behavior is highly sensitive to local conditions.

To address these challenges, this study proposes a Bayesian neural network (BNN)-based framework for RANS turbulence model corrections, extending BNNs beyond uncertainty quantification~\cite{pash2025priori} to correction terms in separated flows. A tensor-based BNN estimates corrections with associated uncertainties, guided by a classifier that isolates shear-dominated regions for targeted input selection. This approach ensures physically consistent corrections, avoids unnecessary modifications, and propagates uncertainties into flow solutions, enhancing RANS prediction reliability across different geometries. The remainder of this paper is structured as follows: Section~\ref{sec:methodology} presents the methodology, detailing the turbulence model corrections, the BNN formulation, and the classifier strategy. Section~\ref{sec:results} discusses the results, focusing on production and Reynolds-stress corrections. Section~\ref{sec:conclusion} summarizes the key findings and outlines potential future research directions.

\section{Methodology}\label{sec:methodology}

\subsection{Bayesian neural networks}

BNNs extend standard neural networks by placing probability distributions over their weights, enabling uncertainty quantification and incorporation of prior knowledge. Instead of optimizing a single set of weights, BNNs infer a posterior distribution using Bayes’ theorem, where the posterior over the weights \( p(w \mid D) \) is proportional to the product of the likelihood \( p(D \mid w) \) and the prior \( p(w) \). Since direct computation is intractable, variational inference approximates the posterior by minimizing the Kullback–Leibler (KL) divergence between an approximate and true posterior,  

\begin{equation}
\mathrm{KL}\!\left(q(w \mid \theta)\, \| \, p(w \mid D)\right)
    = \int q(w \mid \theta)\, \log\!\left(\frac{q(w \mid \theta)}{p(w \mid D)}\right) \, dw ,
\label{eq:kl}
\end{equation}

as shown in Eq.~\eqref{eq:kl}.

The choice of prior significantly impacts model performance by constraining weight distributions, influencing generalization and uncertainty estimation. While isotropic Gaussian priors are common, alternative approaches such as warm-start methods, Gaussian process initialization, and trainable priors offer improved flexibility \cite{pash2025priori}.  
This study employs a trainable prior, similar to KL annealing, where the KL term is dynamically adjusted to balance prior influence and data fitting.

To make optimization tractable, the KL minimization problem is reformulated into the evidence lower bound (ELBO):

\begin{equation}
\mathrm{ELBO}
= \mathbb{E}_{q(w \mid \theta)}\!\left[ \log p(D \mid w) \right]
- \mathrm{KL}\!\left(q(w \mid \theta) \,\|\, p(w)\right),
\label{eq:elbo}
\end{equation}

as shown in Eq.~\eqref{eq:elbo}. Optimizing the ELBO enables BNNs to learn weight distributions while providing uncertainty-aware predictions.

\subsection{Turbulence model correction}

The shear stress transport (SST) \( k-\omega \) model is widely used for turbulence modeling due to its ability to blend the strengths of the \( k-\varepsilon \) and \( k-\omega \) models. However, its reliance on Boussinesq’s assumption, which models the Reynolds stress tensor as proportional to the mean strain rate, introduces errors—particularly in separated flows where turbulence anisotropy plays a significant role.

The Reynolds stress tensor \( \tau_{ij} \) can be decomposed into isotropic and anisotropic components:

\begin{equation}
\tau_{ij} = 2k\left( b_{ij} + \tfrac{1}{3}\,\delta_{ij} \right),
\label{eq:tau}
\end{equation}

as shown in Eq.~\eqref{eq:tau}, where  
\( b_{ij} = -\nu_t / k \, S_{ij} \) is the normalized anisotropic tensor,  
\( \nu_t \) is the eddy viscosity,  
\( k \) is the turbulent kinetic energy (TKE), and \( \delta_{ij} \) is the Kronecker delta.
To address these limitations, this study augments the baseline SST model by incorporating two correction terms:
\( b_{ij}^{\Delta} \), which modifies \( b_{ij} \), and\( P_k^{\Delta} \), which adjusts TKE production \( P_k \)

The governing equations of the augmented SST model are expressed as:

\begin{equation}
\frac{Dk}{Dt}
  = \nabla \cdot (\nu_t \nabla k)
    + P_k + P_k^{\Delta}
    - \beta^{*}\, \omega k,
\qquad
\frac{D\omega}{Dt}
  = \nabla \cdot (\nu_t \nabla \omega)
    + \gamma\,(P_k + P_k^{\Delta})/\nu_t
    - \beta\, \omega^{2},
\label{eq:sst}
\end{equation}

as shown in Eq.~\eqref{eq:sst}, where \( \beta \) and \( \beta^{*} \) are model parameters.  
For derivation and implementation details, refer to \cite{menter1993zonal}.
The correction terms \( b_{ij}^{\Delta} \) and \( P_k^{\Delta} \) are computed using the $k$-corrective frozen RANS approach \cite{schmelzer2020discovery}.  In this approach, values of \( k \), \( U \), and \( b_{ij} \) are extracted from high-fidelity (HF) data, and the \( \omega \)-equation is solved iteratively while freezing other variables to compute deviations in the anisotropic stress tensor and TKE production.

Following the Cayley--Hamilton theorem, $b_{ij}^{\Delta}$ is formulated using a tensor basis expansion, providing a more flexible representation of turbulence anisotropy. Similarly, the correction term $P_{k}^{\Delta}$ follows an analogous formulation, where the tensor basis functions $(T_{ij}^{(n)})$ and invariants $(I_m)$ of the strain rate $(S_{ij})$ and rotation rate $(\omega_{ij})$ tensors define their structure. This correction term also includes an additional term in which $\varepsilon = \omega k$ and $\alpha_4$ is a function of the invariants.

\begin{equation}
b_{ij}^{\Delta} = \sum_{n=1}^{N} T_{ij}^{(n)} \alpha_n(I_m),
\qquad
P_{k}^{\Delta} = 2k\, b_{ij}^{\Delta} \frac{\partial U_i}{\partial x_j} + \alpha_4 \varepsilon ,
\tag{5}
\end{equation}

By incorporating $b_{ij}^{\Delta}$ and $P_{k}^{\Delta}$, the model enhances turbulence representation in regimes where conventional closures lack accuracy as shown in equation (5).

\subsection{Zonal classification}

A key aspect of the proposed methodology is the identification of regions where turbulence model corrections are necessary. Instead of applying corrections globally, a classification method is introduced to selectively modify the turbulence closure in flow regions dominated by shear effects. To achieve this, we adopt the relative importance term analysis (RITA) classifier \cite{luacuatucs2024improving}, inspired by recent clustering approaches \cite{callaham2021learning, srivastava2024generalizably} used to identify dominant physical processes in turbulent flows. The classification is based on the transport equation of TKE, where the relative contributions of production $(P_k)$, dissipation $(D_k)$, convection $(C_k)$, and diffusion $(D_{f,k})$ terms determine the turbulence dynamics at each point in the domain. RITA classifier evaluates the importance of these terms by computing relative ratios

\begin{equation}
\mathrm{RITA}_{P_k/D_k} = \frac{|D_k|}{(|P_k| + |D_k|)}, 
\qquad
TI = \frac{k}{(k + 0.5\,\|U\|^{2})},
\tag{6}
\end{equation}

as demonstrated in equation (6). Cells satisfying the conditions $\mathrm{RITA}_{P_k/D_k} < 0.55$ and $TI \ge 0.12$ are classified as part of the shear zone, ensuring that corrections are only applied where turbulence anisotropy and production imbalances are significant. The classifier assigns a binary variable $\sigma^{*}$, which determines whether the correction terms $b_{ij}^{\Delta}$ and $P_k^{\Delta}$ are activated. This targeted approach prevents unnecessary modifications in well-predicted regions, enhancing the efficiency and stability of the model.

\section{Results and Discussion}\label{sec:results}

\subsection{BNN structure and UQ framework}

The BNN structure follows the tensor basis formulation, ensuring Galilean invariance and frame-independence by leveraging tensor basis functions and invariants. The law of total variance \cite{saltelli2010variance} allows decomposition of the overall predictive uncertainty into aleatoric and epistemic components.

\begin{equation}
\mathrm{Var}(y) 
= \mathbb{E}_{q(w \mid \theta)}\!\left[\mathrm{Var}(y \mid x, D)\right]
+ \mathrm{Var}\!\left(\mathbb{E}_{q(w \mid \theta)}[y \mid x, D]\right),
\tag{7}
\end{equation}

The first term in equation (7) represents aleatoric uncertainty, computed as the mean of the standard deviations across all Monte Carlo (MC) samples, capturing inherent noise in the data. The second term corresponds to epistemic uncertainty, estimated from the variability of the model’s mean predictions, reflecting uncertainty due to limited training data or model limitations. This decomposition ensures a clear distinction between data-driven noise and model uncertainty, aiding in assessing the reliability of turbulence model corrections.

\begin{equation}
\sigma^{2}_{\Delta b_{ij}} = \sum_{n=1}^{N} \left(T^{(n)}_{ij}\right)^{2} \sigma_{n}^{2}(I_m),
\qquad
\sigma^{2}_{P_{k}^{\Delta}} 
= 4k^{2} \sum_{n=1}^{N} \left(T^{(n)}_{ij}\, \partial_j U_i\right)^{2}
\sigma_{n}^{2}(I_m) + \sigma_{\alpha_4}^{2}\, \varepsilon^{2},
\tag{8}
\end{equation}

The variance predictions for aleatoric uncertainty follow equation (8). Regions with high variance indicate areas where the model is uncertain, reflecting sensitivity to missing physics or limited training data. To optimize the BNN parameters, a modified ELBO objective function is used applying MC approximation:

\begin{equation}
\mathrm{ELBO} \approx 
\frac{1}{N} \sum_{i}
\left[ 
\log q(w^{(i)} \mid \theta)
- \log p(w^{(i)})
- \log p(D \mid w^{(i)})
\right],
\tag{9}
\end{equation}
shown in equation (9), where the right hand side terms are the log-probability of approximate variational posterior, prior distribution on weights and likelihood of dataset given the weight of sample, respectively. The BNN loss function is defined as equation (10).

\begin{equation}
\mathrm{Loss}
= \lambda_{1}\mathrm{MSE}
+ \lambda_{2}\log(\sigma^{2})
+ \frac{1}{2\sigma^{2}}
+ \lambda_{3}\mathrm{KL},
\tag{10}
\end{equation}

where MSE enforces accuracy with HF data, $\log(\sigma^{2})$ and $1/2\sigma^{2}$ regulate aleatoric uncertainty and KL divergence controls weight regularization. Hyperparameters, determined through a sensitivity study, are set to $\lambda_{1}=10$, $\lambda_{2}=0.1$, and $\lambda_{3}=0.1$.

\begin{figure}[ht]
    \centering
    \includegraphics[width=0.9\linewidth]{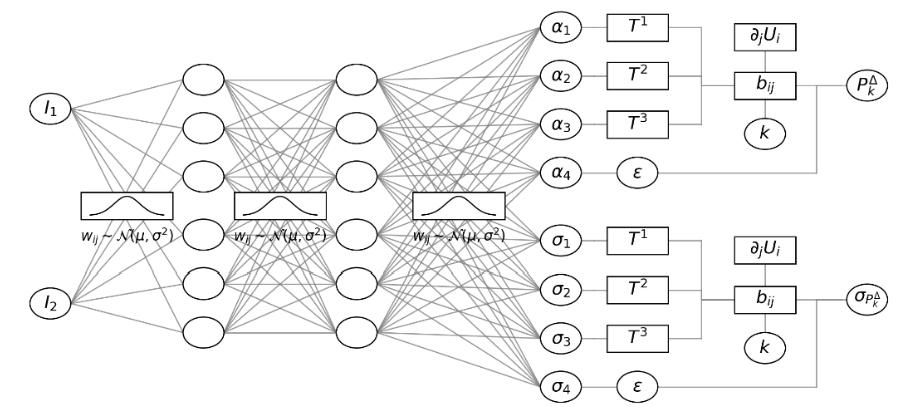}
    \caption{Tensor-basis BNN structure for TKE production correction}
    \label{fig:figure1}
\end{figure}

The BNN framework, illustrated in Figure~\ref{fig:figure1}, is designed to predict the turbulence correction terms $P_{k}^{\Delta}$ and $b_{ij}^{\Delta}$ while quantifying the associated uncertainties. The network architecture for $P_{k}^{\Delta}$ follows the structure shown in the figure, while the architecture for $b_{ij}^{\Delta}$ remains identical except for the output layer, where it predicts only the three non-repetitive independent components of the anisotropic stress tensor.

To ensure optimal feature selection, Mutual Information (MI) ranking is applied, retaining only features with an MI score above 0.8, ensuring informative yet non-redundant inputs. The selected features and tensor basis terms for each correction field are summarized in Table~\ref{tab:table1}.

\begin{table}[t]
\centering
\caption{BNN framework and selected features}
\label{tab:table1}
\begin{tabular}{ll}
\hline
\textbf{Category} & \textbf{Details} \\
\hline
Model architecture & 4 hidden layer, 25 nodes per layer \\
Activation function & Sigmoid \\
Prior distribution & Trainable prior $(\mu, \sigma)$ \\
Dataset & Total samples: 2556, training data: 70\%, validation data: 30\% \\
Training parameters & Epochs: 500, MC Samples: 100 \\
Selected invariants 
& $P_{k}^{\Delta} : I_{1}, I_{2}, R_{e_{s}}, RITA_{P_k/D_k}, RITA_{C_k/D_k}$ \\
& $b_{ij}^{\Delta} : I_{1}, I_{2}, I_{5}, 1 - I_{1}, R_{e_{\Omega}}, RITA_{P_k/D_k}, RITA_{C_k/D_k}$ \\
Tensor basis 
& $P_{k}^{\Delta} : T_{1}, T_{2}, T_{3}$ \\
& $b_{ij}^{\Delta} : T_{1}, T_{2}, T_{3}, T_{4}$ \\
\hline
\end{tabular}
\end{table}

\subsection{Inference}

\begin{figure}[t]
    \centering
    \includegraphics[width=0.95\linewidth]{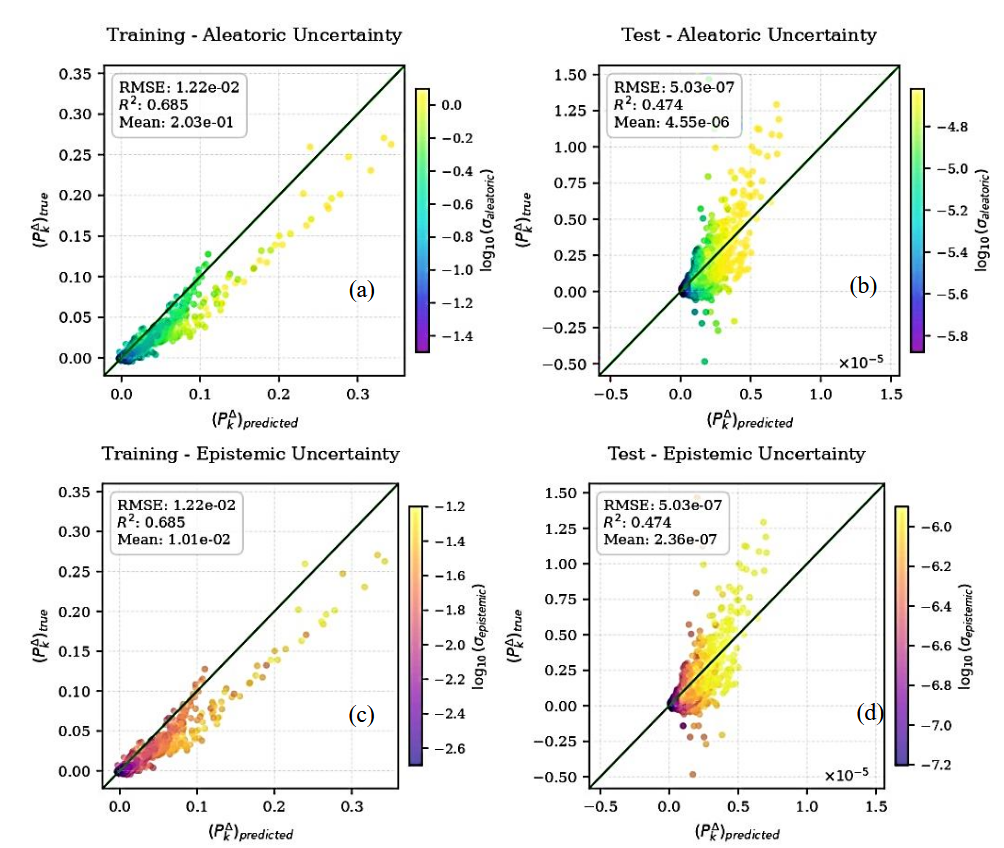}
    \caption{Error metrics for training and test data with uncertainty values for TKE production: training data with epistemic uncertainty (a), training data with aleatoric uncertainty (b), test data with epistemic uncertainty (c) and test data with aleatoric uncertainty (d).}
    \label{fig:figure2}
\end{figure}

The proposed BNN framework is trained and validated using three benchmark turbulent flow cases: the periodic hill with $Re = 10595$ based on hill height \cite{breuer2009flow}, and the curved backward-facing step at $Re = 13700$ based on step height \cite{bentaleb2012large}. To evaluate the generalization capability of the model, a set of parameterized periodic hill cases \cite{xiao2020flows} is used as test scenarios. To evaluate the BNN's inference performance, Figure~\ref{fig:figure2} shows predictions for TKE production correction terms. For training cases, predictions align closely with the one-to-one reference line, indicating high accuracy. Low epistemic uncertainty reflects strong model confidence within familiar data regions, while higher aleatoric uncertainty highlights the inherent variability of turbulent flows. Despite a narrower target range in test cases, the BNN maintains satisfactory predictive accuracy, demonstrating robust generalization to unseen data.

\subsection{Propagation}

To evaluate the effectiveness of the proposed correction strategy, correction terms were tested on both training and test cases by propagating them into OpenFOAM simulations. This validation step ensured that the corrections accurately represent the missing physics and verified the robustness of the methodology across different flow scenarios.

\begin{figure}[t]
    \centering
    \includegraphics[width=0.95\linewidth]{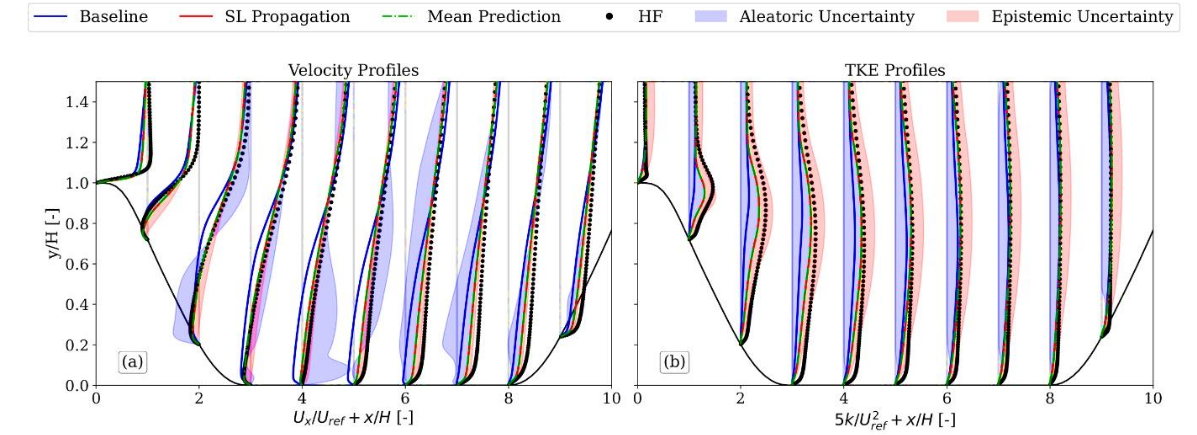}
    \caption{Normalized velocity (a) and TKE (b) profiles for one of the test cases, showing baseline predictions, high-fidelity (HF) data, shear layer (SL) correction propagation, and BNN mean prediction propagation results, along with aleatoric and epistemic uncertainty bounds.}
    \label{fig:figure3}
\end{figure}

Figure~\ref{fig:figure3} illustrates the propagation results, showcasing both the accuracy of mean predictions and associated uncertainties. The BNN predictions, representing the mean of multiple MC samples, reveal two key findings. First, applying correction terms significantly improves predictions, particularly in separated flow regions and highly sheared areas where baseline RANS models typically underperform. This physically manifests as improved prediction of separation bubble size and wake recovery rates. The turbulent production correction, $P_{k}^{\Delta}$, addresses the energetic imbalance in separated regions by enhancing turbulent mixing across high-shear zones, promoting faster momentum exchange between the separation bubble and freestream flow. Importantly, the RITA classifier ensures corrections are applied only in separation regions while preserving the baseline model's accurate boundary layer predictions. Second, the close agreement between BNN-predicted propagation results and those derived from HF data confirms the BNN's ability to capture essential corrections.

Additionally, aleatoric and epistemic uncertainty bounds are presented, with aleatoric uncertainty dominating the predictions. This suggests data sparsity as the primary source of uncertainty. Notably, higher uncertainty is observed near reattachment points and regions with strong velocity gradients, where complex turbulence dynamics pose greater modeling challenges. While the corrections result in some elevated TKE levels compared to HF data, the improved mean flow predictions demonstrate how
they effectively capture the dominant non-equilibrium physics in separated regions, where production-to-dissipation ratios deviate significantly from equilibrium assumptions.

\section{Conclusion}\label{sec:conclusion}

This study proposed a BNN-based framework to improve RANS turbulence models in separated flow regimes, where traditional closures often fail. Using a tensor-basis formulation, the BNN provided Galilean-invariant, uncertainty-aware corrections for Reynolds stress anisotropy and TKE production. A RITA-based classifier enhanced model stability by applying corrections selectively in shear-dominated regions. Validated on periodic hill, and curved backward-facing step cases, with parameterized periodic hills for testing, the framework showed significant improvements in velocity with a systematic overprediction in TKE. The BNN accurately reproduced correction terms, aligning closely with HF data. The physical impact of these corrections is most evident in separation bubbles and wake regions, where modified momentum transport and enhanced mixing capture flow physics that linear eddy-viscosity models miss. While uncertainty quantification revealed higher aleatoric uncertainty compared to epistemic uncertainty, suggesting limitations in training data diversity rather than model structure, these bounds warrant careful interpretation in complex flow regions. This approach demonstrates an efficient integration of machine learning with physics-based constraints, offering improved predictive accuracy and uncertainty quantification. Future work could enhance generalizability by incorporating a broader range of flow conditions for even 3D cases.

\section*{Acknowledgment}
A.E. acknowledges support from the GO-VIKING project (Grant No. 101060826) under the Euratom Research and Training Program.

\section*{References}
\bibliographystyle{ieeetr} 
\bibliography{main}

\end{document}